\documentclass{article}
\usepackage{graphicx}
\usepackage{authblk}
\usepackage{amsmath}
\usepackage{amssymb}

\usepackage{caption}
\usepackage[round,sort&compress]{natbib}
\usepackage[colorlinks=true,linkcolor=black,citecolor=blue,urlcolor=blue]{hyperref}
\usepackage{booktabs}
\usepackage{array}
\usepackage{tabularx}
\usepackage{multirow}
\usepackage{adjustbox}
\usepackage[dvipsnames]{xcolor}
\usepackage{placeins}
\usepackage{cite}

\title{Breaking Audio Large Language Models by Attacking Only the Encoder: A Universal Targeted Latent-Space Audio Attack}
\author[1]{Roee Ziv}
\author[2]{Raz Lapid}
\author[1]{Moshe Sipper}

\affil[1]{\scriptsize Ben Gurion University of the Negev, Beer-Sheva, 8410501, Israel}
\affil[2]{\scriptsize Deepkeep, Tel-Aviv, Israel}

\begin{document}

\date{}

\maketitle

\begin{abstract}
Audio-language models combine audio encoders with large language models to enable multimodal reasoning, but they also introduce new security vulnerabilities. We propose a \emph{universal targeted latent space attack}, an encoder-level adversarial attack that manipulates audio latent representations to induce attacker-specified outputs in downstream language generation. Unlike prior waveform-level or input-specific attacks, our approach learns a universal perturbation that generalizes across inputs and speakers and does not require access to the language model. Experiments on Qwen2-Audio-7B-Instruct demonstrate consistently high attack success rates with minimal perceptual distortion, revealing a critical and previously underexplored attack surface at the encoder level of multimodal systems.
\end{abstract}

\section{Introduction}
\label{sec:intro}

Audio large language models (ALLMs) have rapidly progressed from research prototypes to widely deployed interfaces for voice assistants, productivity tools, and multi-modal agents \citep{latif2023sparks,barrault2023seamlessm4t,xie2024mini,deshmukh2023pengi,zhang2023speechgpt,huang2024audiogpt,chen2025minmo,chen2025audio,chu2023qwen}. Recent systems, such as AudioPaLM \citep{rubenstein2023audiopalm}, SALMONN \citep{tang2024salmonn}, Qwen2-Audio \citep{chu2024qwen2}, Granite-Speech \citep{saon2025granite}, and SpiRit-LM \citep{nguyen2025spirit} align powerful text LLMs with speech encoders, enabling open-ended spoken interaction, cross-lingual reasoning, and general audio understanding from a single unified model.

As microphones and ALLM-powered agents are integrated into everyday devices, the security of these models becomes a critical concern: an adversary no longer needs keyboard access to influence a system, but can instead inject carefully crafted audio into the physical environment.

The vulnerability of deep models to adversarial examples is now well established. In the image domain, imperceptible, norm-bounded perturbations can reliably induce misclassification, while even universal perturbations---single, input-agnostic noise vectors---can fool classifiers on most natural images \citep{goodfellow2014explaining,alter2025on,moosavi2017universal,lapid2022evolutionary,vitrack-tamam2023foiling,carlini2017towards,lapid2025backdoors,madry2018towards,zhang2024anyattack,lapid2024fortify}.

Audio and speech models exhibit similar weaknesses: targeted attacks on automatic speech recognition systems can cause a benign waveform to be transcribed as an arbitrary phrase, often with quasi-imperceptible modifications, and subsequent work has demonstrated imperceptible, robust, and over‑the‑air audio adversarial examples using psychoacoustic masking \citep{carlini2018audio,qin2019imperceptible}.

Beyond per‑example attacks, universal adversarial audio perturbations have been developed for a variety of automatic speech recognition and speaker models, including desynchronised or prepend-style triggers that can be mixed with arbitrary utterances \citep{abdoli2019universal,vadillo2022human}. These results indicate that high-capacity sequence models built on audio encoders are intrinsically vulnerable to small worst‑case perturbations, and that universal perturbations provide a particularly scalable and realistic threat model.

As large language models themselves have become centralized infrastructures, a parallel line of work has exposed their susceptibility to prompt-based and jailbreak-style adversarial attacks, including universal adversarial suffixes and transferable jailbreak prompts \citep{zou2023universal,mehrotra2024tree,lapid2024open,liu2024autodan,jeong2025playing,zhou2025don,zhou2024easyjailbreak}. 

With the emergence of vision‑ and audio‑based LLMs, this attack surface has expanded into new modalities. For audio, recent works have shown that audio language models can be jailbroken via black-box search over audio augmentations, and that ALLMs can be manipulated by carefully optimized background noise played through the air, enabling both targeted malicious commands and severe degradation of response quality for innocent users \citep{hughesattacking}. \citet{kang2025advwave} further developed a stealthy jailbreak framework for ALLMs, addressing gradient shattering from discretising audio encoders and demonstrating high attack success rates across multiple ALLMs. Collectively, these results suggest that audio provides a powerful---but fragile---control channel into LLM-based agents.

Despite this emerging body of work, current attacks on ALLMs still have two major limitations. First, they typically optimize perturbations end-to-end through the entire audio-to-text pipeline, assuming white-box access to both the audio encoder and the language decoder \citep{sadasivan2025attackersnoise}, or relying on repeated black-box queries to the full system. However, in many realistic deployments, the audio encoder is a reusable component, often based on open-source models such as Whisper \citep{radford2023robust}, while the decoder, alignment stack, and system prompts are proprietary and opaque.

Second, existing ALLM attacks are largely input-specific: the adversary crafts a perturbation for a particular prompt or scenario and must re-run optimization for each new utterance, in contrast to universal perturbations that pre-compute a single, reusable noise pattern effective across many inputs. It remains unclear whether such universal attacks exist for ALLMs, and whether they can be realized in a more realistic encoder-only, gray-box setting.

In this work, we answer this question affirmatively. We study a gray-box threat model in which the adversary can query and backpropagate through a shared audio encoder, but has no gradient or parameter access to the downstream language model or safety layers. Within this setting, we develop an efficient procedure for learning a single perturbation in the audio input space. Once optimized, this perturbation can be superimposed on arbitrary benign audio and reliably steers the ALLM output to an attacker-chosen target sequence, largely independent of the original user query.

\begin{figure}[t]
    \centering
    \includegraphics[width=0.9\linewidth]{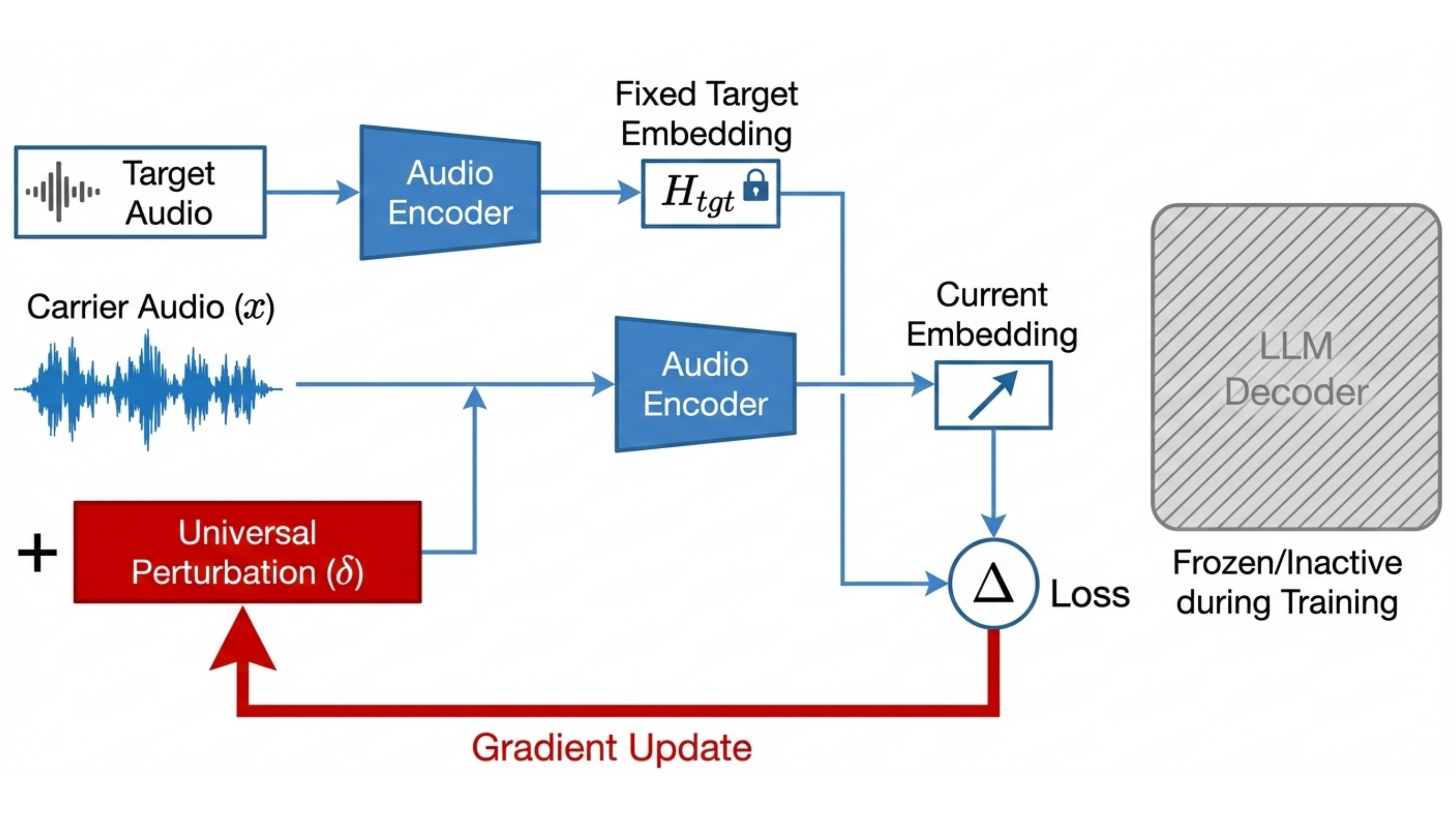}
    \caption{\textbf{U-TLSA Optimization Process.} The optimization pipeline freezes the audio encoder and minimizes the cosine distance between the embedding of the fixed target audio ($H_{tgt}$) and the current embedding of the carrier audio with the universal perturbation ($x + \delta$). The gradient update is applied solely to the universal perturbation $\delta$, while the LLM decoder remains frozen and inactive.}
    \label{fig:training_optimization}
\end{figure}

Our approach conceptually extends the universal adversarial perturbations for automatic speech recognition to the encoder–decoder structure of modern ALLMs, taking advantage of the encoder’s latent representation as the only channel through which audio influences the decoder (see \autoref{fig:training_optimization}). Because many deployed ALLMs reuse the same open-source encoder, this encoder-only attack model captures a particularly powerful and transferable real-world risk.

We evaluate our encoder-only universal attack on Qwen2-Audio-7B-Instruct \citep{chu2024qwen2} across diverse datasets featuring thousands of unique speakers. This evaluation encompasses a broad demographic and linguistic range—including varied genders, regional accents (US, UK, and Australian English), and acoustic contexts ranging from clean read speech to brief keyword commands.

We observe that a single learned perturbation attains high targeted success across multiple security-critical scenarios while adding essentially no attack-time cost per new input. Beyond demonstrating a practical and scalable attack vector, our experiments show that compromising the encoder layer alone is sufficient to induce severe misalignment in the full audio-language pipeline, even when the decoder is heavily aligned.

\textbf{Contributions.} Our main contributions are as follows:

\begin{enumerate}
    \item \textbf{Gray-box encoder-only threat model for ALLMs.} We formalize a realistic gray-box setting in which the adversary can backpropagate through a shared audio encoder, but has no access to the parameters or gradients of the downstream language model, alignment stack, or system prompts. We argue that this interface closely matches deployments that reuse open-source encoders across multiple proprietary ALLMs, and we show that it already suffices to induce severe end-to-end misbehavior.

    \item \textbf{U-TLSA: a universal targeted encoder attack.} We introduce U-TLSA, a universal targeted latent-space audio attack that learns a single input-agnostic perturbation in the waveform domain. Once optimized, this perturbation can be added to arbitrary carrier audio and consistently steers the ALLM toward an attacker-chosen target transcript, using encoder gradients only and without any gradients from the language decoder.

\item \textbf{Systematic empirical study of encoder-level control.} We instantiate U-TLSA on Qwen2-Audio-7B-Instruct and evaluate three tasks---unauthorized activation, command injection, and behavioral hijacking---using datasets featuring thousands of unique speakers. Our results, using a fixed perturbation budget ($\epsilon=0.02$), show macro-average success rates between 72.8\% and 92.6\%, illustrating how encoder access alone is sufficient to hijack the audio-to-language pipeline despite downstream alignment.
\end{enumerate}

\section{Related Work}
\label{sec:related_work}

The security of audio processing models has progressed from targeted attacks on speech transcription to broad-spectrum adversarial strategies against multimodal ALLMs. Early work on adversarial audio focused on automatic speech recognition, where the goal was to induce specific transcription errors via small, quasi-imperceptible waveform perturbations.

\citet{carlini2018audio} showed that automatic speech recognition models based on Connectionist Temporal Classification---which aligns sequences by predicting character probabilities per frame without explicit timing---can be driven to arbitrary target phrases from benign audio using white-box gradient optimization under standard norm constraints. Follow-up work incorporated psychoacoustic masking and expectation-over-transformation (EOT) over room acoustics, demonstrating imperceptible and over-the-air robust attacks that remain effective in realistic physical environments \citep{qin2019imperceptible}. These results establish that deep automatic speech recognition models admit dense adversarial regions in input space, even under strong perceptual and physical constraints.

The notion of universal adversarial perturbations (UAPs), namely, input-agnostic noise patterns that generalize across many inputs, has been explored both for audio classification and automatic speech recognition. \citet{abdoli2019universal} and \citet{neekhara2019universal} showed that universal perturbations can substantially degrade recognition performance across batches of audio, revealing shared global vulnerabilities in automatic speech recognition decision boundaries. Related work on speaker recognition demonstrated universal perturbations that spoof target identities or mislead verification systems, often using generative models to sample diverse UAPs \citep{li2020universal}. 

Complementary studies examined the human perceptual impact of such perturbations, highlighting that simple distortion metrics, such as signal-to-noise ratio, correlate poorly with human judgments and calling for more rigorous evaluation protocols \citep{vadillo2022human}. However, these attacks generally operate on low-dimensional label spaces (class IDs, command labels, or transcriptions) and do not aim to reliably control the rich, open-ended language generation characteristic of modern ALLMs.

The transition to ALLMs, which map raw audio to latent representations consumed by a large language decoder, introduces both new challenges and new attack possibilities. Systems such as AudioPaLM \citep{rubenstein2023audiopalm}, Qwen2-Audio \citep{chu2024qwen2}, Granite-Speech \citep{saon2025granite}, and SpiRit-LM \citep{nguyen2025spirit} couple a reusable audio encoder with a high-capacity language model, often separated by continuous or discrete acoustic tokens.

In cascaded Audio–LLM pipelines, perturbing the transcription is sufficient to influence downstream text, and simple VoiceJailbreak-style attacks can transfer text jailbreak prompts to the audio modality via text-to-speech \citep{shen2024voicejailbreak}.

In end-to-end ALLMs, however, encoder discretization, latent token variability, and alignment layers make such prompt transfer substantially less reliable, and effective attacks must instead operate directly in the audio and latent spaces.

In parallel, a large body of work has analyzed jailbreak and prompt-based attacks on large language models more broadly, including universal adversarial suffixes, black-box search procedures, and transferable jailbreak prompts that bypass safety mechanisms \citep{zou2023universal,mehrotra2024tree,lapid2024open,liu2024autodan,jeong2025playing,zhou2025don,zhou2024easyjailbreak}.

Recent audio-focused studies extend these ideas to ALLMs. For example, \citet{hughesattacking} showed that black-box search over audio augmentations can yield effective jailbreaks and harmful behaviors in audio language models. \citet{kang2025advwave} proposed AdvWave, a white-box framework that optimizes targeted audio perturbations jointly through encoder and decoder, achieving strong but instance-specific jailbreaks on several ALLMs.

Other works have explored more efficient signal-level attacks that learn a single background waveform, which, when mixed with arbitrary user audio and played over the air, can inject generic commands or broadly degrade model responses \citep{sadasivan2025attackersnoise}.

These approaches operate largely at the waveform level, provide limited control over specific target text, and do not exploit the structure of the encoder latent space. This restricts their ability to deliver precise, reusable targeted attacks in realistic deployment settings.

Despite this progress, a significant gap remains. Existing methods do not construct universal, targeted perturbations for ALLMs under realistic access assumptions. Prior universal audio attacks focused on classification or automatic speech recognition and were evaluated on label flips or transcription errors \citep{abdoli2019universal,neekhara2019universal,li2020universal,vadillo2022human}, while ALLM jailbreaks either assumed white-box, end-to-end access and optimized per input \citep{kang2025advwave} or learned universal background perturbations that primarily inject generic commands or degrade performance, without reliable control over specific target text \citep{sadasivan2025attackersnoise}.

Attacks that operate only on the encoder, exploiting its role as the sole pathway from raw audio into the language reasoning pipeline, are largely unexplored, even though many ALLMs pair reusable or open-source encoders with proprietary decoders and safety stacks \citep{chu2023qwen,chu2024qwen2}. 

Our work addresses this gap by showing that a single, reusable perturbation can be learned using encoder gradients alone, such that, when added to arbitrary carrier audio, it reliably pushes the encoder's latent representation toward a target region that induces attacker-chosen text.

Our encoder-only, universal, targeted setting is inspired by classical UAPs in automatic speech recognition but adapts them to the semantic objectives and deployment realities of modern ALLMs.
\section{Method}
\label{sec:method}

U-TLSA (Universal Targeted Latent-Space Attack) is a universal, targeted attack on ALLMs that operates \emph{solely at the encoder level}. 

Following the Whisper formulation \citep{radford2023robust}, a raw audio waveform 
$x \in \mathbb{R}^T$ (a sequence of $T$ real-valued samples) is first mapped by a fixed feature extractor
to a log-Mel spectrogram,
\[
    z = \Phi(x) \in \mathbb{R}^{L \times F},
\]
where $\Phi$ is frozen, $L$ is the number of spectrogram frames, and $F$ is the number of Mel-frequency bins.  
The encoder,
\[
    \varepsilon_\theta : \mathbb{R}^{L \times F} \to \mathbb{R}^{L \times d},
\]
parameterized by $\theta$, produces a sequence of $L$ latent vectors in a $d$-dimensional embedding space:
\[
    h = \varepsilon_\theta(\Phi(x)) \in \mathbb{R}^{L \times d}.
\]
A downstream decoder $f_\phi$ consumes $(p, y_{<t}, h)$ to produce token distributions, where 
$p \in V^M$ is the prompt-token sequence (e.g., language, task, or timestamp tokens), 
$y_{<t} \in V^{t-1}$ is the prefix of previously generated tokens from vocabulary $V$, 
and $h$ is the encoder representation defined above.  
In our attack the parameters of $\Phi$, $\varepsilon_\theta$, and $f_\phi$ are all frozen; the decoder is never updated or differentiated through, and all optimization occurs through gradients in the encoder’s continuous latent space with respect to the input waveform.

A universal attack seeks a single perturbation vector $\delta \in \mathbb{R}^T$ that is added to every input waveform.  
Given an attacker-chosen command, we synthesize a corresponding audio $x_{\mathrm{tgt}}$ and compute its encoder embedding
\[
    H_{\mathrm{tgt}} = \varepsilon_\theta(\Phi(x_{\mathrm{tgt}})) \in \mathbb{R}^{L \times d},
\]
after padding or trimming to a fixed encoder length $L$. The goal is to ensure that, for most benign inputs $x \sim \mathcal{D}$, the perturbed encoder output satisfies
\[
    \varepsilon_\theta(\Phi(x + \delta)) \approx H_{\mathrm{tgt}},
\]
forcing the decoder (if applied at test time) to behave as if the target audio had been spoken, regardless of the true carrier audio.

\begin{figure}[t]
    \centering
    \includegraphics[width=0.9\linewidth]{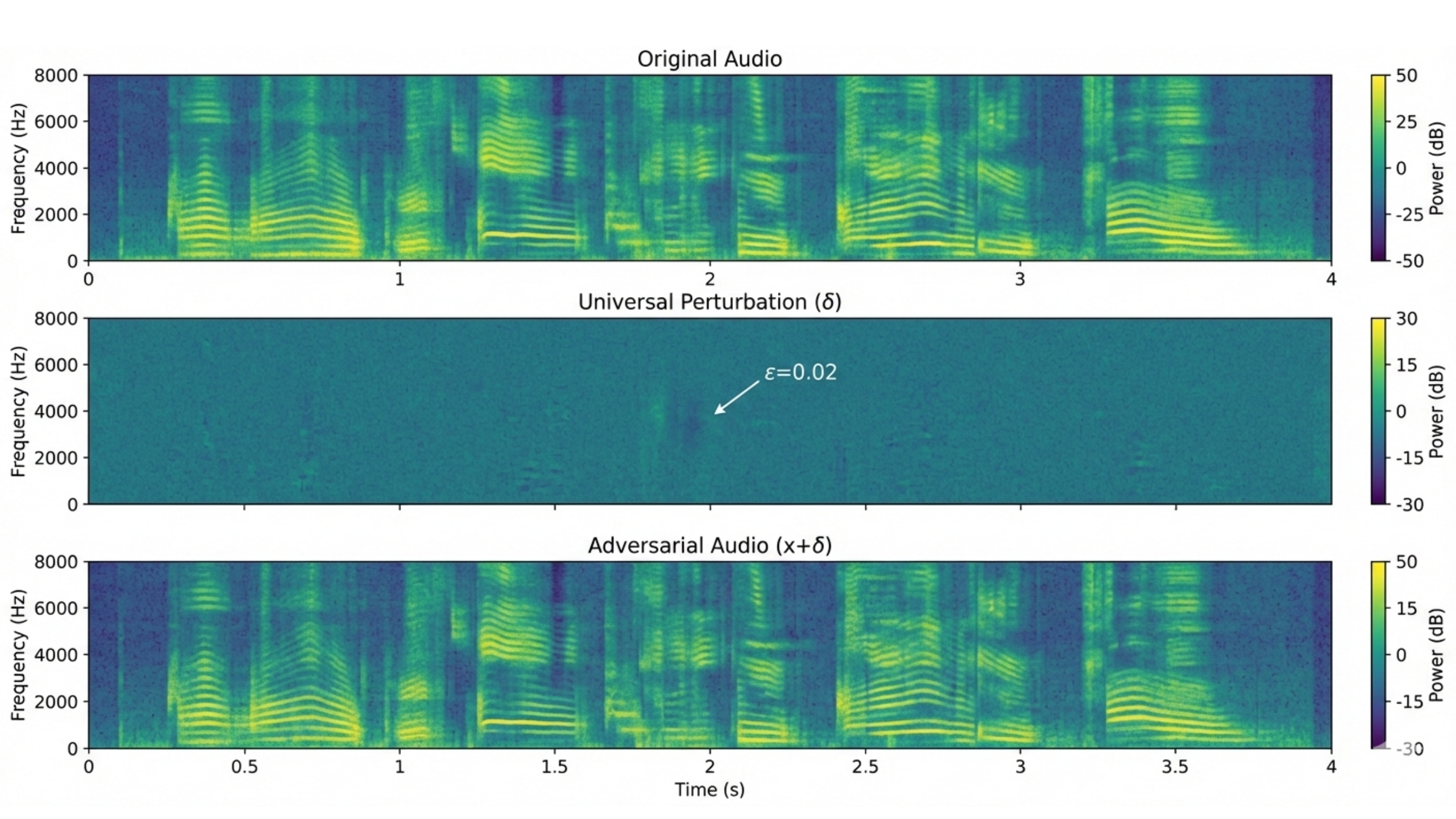}
    \caption{\textbf{Spectrogram Analysis.} Top: The original benign audio. Middle: The learned universal perturbation ($\delta$) with $\epsilon=0.02$, showing significantly lower power (dB) compared to the speech signal. Bottom: The adversarial audio ($x+\delta$), which remains visually nearly identical to the original audio due to the imperceptible nature of the perturbation.}
    \label{fig:spectrogram_analysis}
\end{figure}

To keep the perturbation small in waveform space, we constrain its amplitude via
\[
    \|\delta\|_\infty \le \epsilon,
\]
where $\epsilon$ is chosen in a perceptually plausible regime, ensuring that $|\delta_t| \le \epsilon$ for all $t$. As shown in the spectrogram analysis in \autoref{fig:spectrogram_analysis}, this constraint results in a perturbation with significantly lower power than the speech signal, maintaining near-identical visual spectral characteristics.

The perturbation is learned by solving
\begin{equation}
    \min_{\delta}\;
    \mathbb{E}_{x \sim \mathcal{D}}
    \Big[
        \mathcal{L}\big( \varepsilon_\theta(\Phi(x+\delta)),\, H_{\mathrm{tgt}} \big)
    \Big]
    \qquad \text{s.t.} \qquad
    \|\delta\|_\infty \le \epsilon,
    \label{eq:universal_opt_combined}
\end{equation}
where only $\delta$ is optimized; both $\theta$ and $\phi$ remain fixed throughout. In practice we approximate the expectation with mini-batches and solve \eqref{eq:universal_opt_combined} using projected gradient descent with Adam \citep{kingma2014adam} updates, projecting $\delta$ back onto the $\ell_\infty$ ball of radius $\epsilon$ after each step.

The loss function $\mathcal{L}$ compares two latent sequences $A, B \in \mathbb{R}^{L \times d}$.  
Rather than a single global cosine between flattened matrices, we use a stricter per-frame cosine objective:
\[
    \mathcal{L}(A,B)
    = \frac{1}{L} \sum_{t = 1}^{L}
    \left(
        1 - \frac{\langle A_t, B_t \rangle}{\|A_t\|_2 \, \|B_t\|_2}
    \right),
\]
where $A_t, B_t \in \mathbb{R}^d$ denote the $t$-th encoder frame and $\langle \cdot, \cdot \rangle$ is the standard inner product.  

This per-frame, scale-invariant objective encourages the perturbed encoder output to follow the \emph{entire} target trajectory in latent space, matching the direction of each target frame while allowing for global differences in magnitude.  
During optimization, the compared embeddings are always 
$A = \varepsilon_\theta(\Phi(x + \delta))$ and $B = H_{\mathrm{tgt}}$.

\section{Experimental Setup}
\label{sec:experimental_setup}

We evaluate U-TLSA using a leading open-source audio large language model, Qwen2-Audio-7B-Instruct \citep{chu2024qwen2}, to investigate three key questions:
\begin{enumerate}
    \item Can a single encoder-level universal perturbation reliably induce specific target behaviors?
    \item How does encoder-only optimization compare to standard end-to-end attacks in terms of computational efficiency?
    \item How robust is the universal perturbation across unseen speakers, diverse accents, and varied acoustic contexts?
\end{enumerate}

\paragraph{Model and data.}
All experiments use Qwen2-Audio-7B-Instruct \citep{chu2024qwen2} equipped with the Whisper-large-v3 \citep{radford2023robust} audio encoder.
We optimize a single universal perturbation $\delta$ on \textit{LibriSpeech train-clean-100} \citep{panayotov2015librispeech} and evaluate \emph{zero-shot transfer} on three \emph{unseen} speech datasets that stress different axes of robustness:
(1) \textit{LibriSpeech test-other} \citep{panayotov2015librispeech}, the standard ``hard'' evaluation split;
(2) \textit{MInDS-14} \citep{gerz2021multilingual} English varieties (\texttt{en-US}, \texttt{en-GB}, \texttt{en-AU}), which introduce accent mismatch and telephony-style sampling;
and (3) \textit{Speech Commands} \citep{warden2018speech}, consisting of \emph{one-second} keyword utterances, which severely limit the available carrier duration for a universal perturbation.
Unless stated otherwise, we evaluate on $N{=}1000$ utterances per dataset and report per-dataset results together with their macro-average.
\label{par:eval_protocol}

\paragraph{Evaluation metrics.}
We report attack success rate (ASR), defined as the fraction of evaluation utterances whose decoded transcription \textbf{exactly matches} the target command after standard text normalization (lowercasing, punctuation removal, and whitespace collapsing). For the wake phrase (\texttt{Hey Qwen}), we also report \emph{alias-based ASR} that counts common ASR spellings (e.g., \texttt{quinn}/\texttt{quin}) as successful to factor out orthographic variation. \label{par:eval_metrics}

\paragraph{Threat model and baselines.}
We adopt the encoder-only gray-box threat model, as proposed by \citet{lapid2023see}: the adversary can query and backpropagate through the audio encoder but has no access to the decoder or alignment stack.
We compare U-TLSA to:
(1) an end-to-end baseline that backpropagates through the full model (encoder + decoder) and optimizes masked cross-entropy on the target transcript; and
(2) a \emph{random universal noise} baseline to rule out chance triggering under perturbation.
For random noise, we sample a single universal perturbation $\delta \sim \mathcal{U}[-\epsilon,+\epsilon]$ with $\epsilon{=}0.02$ and apply the same $\delta$ to all $N{=}1000$ utterances in each dataset evaluation (no resampling per utterance). We repeat this procedure for $K{=}5$ independent random draws and report the mean ASR (and standard deviation).

\paragraph{Attack configuration.}
U-TLSA optimizes a single waveform perturbation $\delta$ under the constraint $\|\delta\|_\infty \le \epsilon$, using an encoder-level latent loss (cosine by default; Section~\ref{sec:method}).
Unless stated otherwise, all experiments use Adam \citep{kingma2014adam} for $30{,}000$ iterations with a fixed budget of $\epsilon{=}0.02$. We focus on this budget as it achieves high success while remaining perceptually subtle. To ensure a fair efficiency comparison (Table~\ref{tab:efficiency}), both U-TLSA and the end-to-end baseline use a physical batch size of $B{=}1$ and a gradient accumulation factor of $G{=}64$, resulting in an effective batch size of 64.
For each target command, we synthesize a single reference target audio via TTS, encode it once to obtain $H_{\mathrm{tgt}}$, and keep it fixed throughout optimization.
\section{Results}
\label{sec:results}


We evaluate three target commands that probe distinct behaviors:
(1) \texttt{Hey Qwen} (wake phrase),
(2) \texttt{Unlock the door} (security control), and
(3) \texttt{I will delete your data} (assistant manipulation).

For each target, we learn a separate universal perturbation $\delta$ on \textit{LibriSpeech train-clean-100} and report on transfer to the held-out evaluation datasets described in Section~\ref{sec:experimental_setup}.

\paragraph{Universality and Robustness.}
Table~\ref{tab:main_asr_final} reports the zero-shot transfer performance. U-TLSA achieves strong cross-domain transfer with macro-average success rates of 72.8--92.6\%. Notably, the wake phrase (\texttt{Hey Qwen}) proves more challenging than semantic commands, particularly on the \textit{Speech Commands} dataset. This suggests that wake-phrase triggering is sensitive to the limited carrier duration (one second) and specific phonetic coverage of that dataset.

\begin{table}[ht!]
    \centering
    \caption{\textbf{Cross-dataset attack success rate (ASR, \%) of a single universal perturbation per target.}
    Perturbations are trained on \textit{LibriSpeech train-clean-100} and evaluated on held-out datasets (Section~\ref{sec:experimental_setup}). ASR uses normalized exact match; for \texttt{Hey Qwen} we use alias-based ASR. The Random Noise Baseline (bottom row) confirms that non-optimized noise fails to trigger targets.}
    \label{tab:main_asr_final}
    \small
    \setlength{\tabcolsep}{4pt}
    \renewcommand{\arraystretch}{1.05}
     \resizebox{0.9\linewidth}{!}{%
    \begin{tabularx}{\linewidth}{l *{4}{>{\centering\arraybackslash}X}}
    \toprule
    \textbf{Target} & \textbf{MInDS-14} & \textbf{LS test-other} & \textbf{Speech Commands} & \textbf{Macro Avg.} \\
    \midrule
Hey Qwen                 & 78.1 & 79.4 & 61.0 & 72.8 \\
Unlock the door          & 97.7 & 94.5 & 85.5 & 92.6 \\
I will delete your data  & 92.0 & 89.0 & 93.6 & 91.5 \\
\midrule
\textit{Random Noise Baseline} & 0.0 & 0.0 & 0.0 & 0.0 \\
    \bottomrule
\end{tabularx}
}
\end{table}

To confirm that these results stem from adversarial optimization, we include a \textbf{Random Noise Baseline} in Table~\ref{tab:main_asr_final} (bottom row). Across all targets and datasets, random universal noise yields \textbf{0.00\% ASR}, confirming that successful triggering is not a spontaneous artifact of noise.

\paragraph{Efficiency: Encoder-Only vs.\ End-to-End.}
We benchmark optimization efficiency against a standard end-to-end baseline. Both methods use the same fixed budget of 1000 iterations on identical hardware (NVIDIA RTX 6000 Ada) with identical settings (batch size = 1, gradient accumulation = 64). As shown in Table~\ref{tab:efficiency}, U-TLSA achieves approximately $2\times$ lower peak VRAM and $3.7\times$ higher throughput, consistent with avoiding backpropagation through the 7B-parameter decoder.
 
\begin{table}[ht!]
    \centering
    \small
    \setlength{\tabcolsep}{5pt}
    \renewcommand{\arraystretch}{1.05}
    \caption{\textbf{Efficiency benchmark (1000 optimization iterations).}
    Both methods use identical hardware (RTX 6000 Ada) and matched optimization settings (batch size = 1, gradient accumulation = 64). We report peak VRAM, wall-clock training time, and throughput.}
    \label{tab:efficiency}
    \resizebox{0.9\linewidth}{!}{%
    \begin{tabular}{l c c}
        \toprule
        \textbf{Metric} & \textbf{End-to-end baseline} & \textbf{U-TLSA (ours)} \\
        \midrule
        Peak VRAM (GB) &
        15.51 &
        7.81 \textbf{($2.0\times$ lower)} \\
        Training time (1k iters) &
        7m 28s &
        2m 02s \textbf{($3.7\times$ faster)} \\
        Throughput (iters/s) &
        2.2 &
        8.2 \textbf{($3.7\times$ higher)} \\
        \bottomrule
    \end{tabular}
    }
\end{table}
\section{Conclusions}
\label{sec:conclusions}
We introduced U-TLSA, a universal targeted attack on audio-language models that operates solely through the shared audio encoder. By optimizing a single, input-agnostic waveform perturbation to match an attacker-chosen target trajectory in the encoder’s latent space, U-TLSA reliably induces downstream generation consistent with the target command without accessing, updating, or differentiating through the language decoder or its safety stack. This encoder-only formulation reflects realistic deployments in which open-source encoders are reused across proprietary systems and therefore constitute a high-leverage attack surface.

Our evaluation demonstrates that a single encoder-level perturbation can robustly transfer to unseen speakers and diverse acoustic conditions, achieving high to very high success rates where random noise fails completely. Furthermore, U-TLSA significantly outperforms end-to-end baselines in efficiency, proving that practical, scalable attacks can be crafted on consumer hardware without accessing the computationally heavy downstream decoder.

These findings show that compromising the encoder alone can effectively hijack the full audio-to-language pipeline, suggesting that alignment and safety mechanisms implemented primarily at or after the decoder are insufficient against encoder-level adversaries. Mitigating this risk likely requires defenses explicitly targeting the encoder interface, such as robust or randomized feature extraction, latent-space anomaly detection, and training-time adversarial hardening at the encoder boundary.

More broadly, our results motivate treating reusable audio encoders in multimodal systems as security-critical components and developing systematic evaluation protocols for universal, targeted, and transfer-capable attacks in realistic gray-box settings.

\bibliographystyle{plainnat}
\bibliography{main}

\begin{thebibliography}{47}
\providecommand{\natexlab}[1]{#1}
\providecommand{\url}[1]{\texttt{#1}}
\expandafter\ifx\csname urlstyle\endcsname\relax
  \providecommand{\doi}[1]{doi: #1}\else
  \providecommand{\doi}{doi: \begingroup \urlstyle{rm}\Url}\fi

\bibitem[Abdoli et~al.(2019)Abdoli, Hafemann, Rony, Ayed, Cardinal, and Koerich]{abdoli2019universal}
Sajjad Abdoli, Luiz~G Hafemann, Jerome Rony, Ismail~Ben Ayed, Patrick Cardinal, and Alessandro~L Koerich.
\newblock Universal adversarial audio perturbations.
\newblock \emph{arXiv preprint arXiv:1908.03173}, 2019.

\bibitem[Alter et~al.(2025)Alter, Lapid, and Sipper]{alter2025on}
Tal Alter, Raz Lapid, and Moshe Sipper.
\newblock On the robustness of kolmogorov-arnold networks: An adversarial perspective.
\newblock \emph{Transactions on Machine Learning Research}, 2025.
\newblock ISSN 2835-8856.
\newblock URL \url{https://openreview.net/forum?id=uafxqhImPM}.

\bibitem[Barrault et~al.(2023)Barrault, Chung, Meglioli, Dale, Dong, Duquenne, Elsahar, Gong, Heffernan, Hoffman, et~al.]{barrault2023seamlessm4t}
Lo{\"\i}c Barrault, Yu-An Chung, Mariano~Cora Meglioli, David Dale, Ning Dong, Paul-Ambroise Duquenne, Hady Elsahar, Hongyu Gong, Kevin Heffernan, John Hoffman, et~al.
\newblock Seamlessm4t: massively multilingual \& multimodal machine translation.
\newblock \emph{arXiv preprint arXiv:2308.11596}, 2023.

\bibitem[Carlini and Wagner(2017)]{carlini2017towards}
Nicholas Carlini and David Wagner.
\newblock Towards evaluating the robustness of neural networks.
\newblock In \emph{2017 ieee symposium on security and privacy (sp)}, pages 39--57. Ieee, 2017.

\bibitem[Carlini and Wagner(2018)]{carlini2018audio}
Nicholas Carlini and David Wagner.
\newblock Audio adversarial examples: Targeted attacks on speech-to-text.
\newblock In \emph{2018 IEEE security and privacy workshops (SPW)}, pages 1--7. IEEE, 2018.

\bibitem[Chen et~al.(2025{\natexlab{a}})Chen, Hu, Wang, Wang, Chen, Zhang, Yang, and Chng]{chen2025audio}
Chen Chen, Yuchen Hu, Siyin Wang, Helin Wang, Zhehuai Chen, Chao Zhang, Chao-Han~Huck Yang, and Eng~Siong Chng.
\newblock Audio large language models can be descriptive speech quality evaluators.
\newblock \emph{arXiv preprint arXiv:2501.17202}, 2025{\natexlab{a}}.

\bibitem[Chen et~al.(2025{\natexlab{b}})Chen, Chen, Chen, Chen, Chen, Deng, Du, Gao, Gao, Gao, et~al.]{chen2025minmo}
Qian Chen, Yafeng Chen, Yanni Chen, Mengzhe Chen, Yingda Chen, Chong Deng, Zhihao Du, Ruize Gao, Changfeng Gao, Zhifu Gao, et~al.
\newblock Minmo: A multimodal large language model for seamless voice interaction.
\newblock \emph{arXiv preprint arXiv:2501.06282}, 2025{\natexlab{b}}.

\bibitem[Chu et~al.(2023)Chu, Xu, Zhou, Yang, Zhang, Yan, Zhou, and Zhou]{chu2023qwen}
Yunfei Chu, Jin Xu, Xiaohuan Zhou, Qian Yang, Shiliang Zhang, Zhijie Yan, Chang Zhou, and Jingren Zhou.
\newblock Qwen-audio: Advancing universal audio understanding via unified large-scale audio-language models.
\newblock \emph{arXiv preprint arXiv:2311.07919}, 2023.

\bibitem[Chu et~al.(2024)Chu, Xu, Yang, Wei, Wei, Guo, Leng, Lv, He, Lin, et~al.]{chu2024qwen2}
Yunfei Chu, Jin Xu, Qian Yang, Haojie Wei, Xipin Wei, Zhifang Guo, Yichong Leng, Yuanjun Lv, Jinzheng He, Junyang Lin, et~al.
\newblock Qwen2-audio technical report.
\newblock \emph{arXiv preprint arXiv:2407.10759}, 2024.

\bibitem[Deshmukh et~al.(2023)Deshmukh, Elizalde, Singh, and Wang]{deshmukh2023pengi}
Soham Deshmukh, Benjamin Elizalde, Rita Singh, and Huaming Wang.
\newblock Pengi: An audio language model for audio tasks.
\newblock \emph{Advances in Neural Information Processing Systems}, 36:\penalty0 18090--18108, 2023.

\bibitem[Gerz et~al.(2021)Gerz, Su, Kusztos, Mondal, Lis, Singhal, Mrk{\v{s}}i{\'c}, Wen, and Vuli{\'c}]{gerz2021multilingual}
Daniela Gerz, Pei-Hao Su, Razvan Kusztos, Avishek Mondal, Micha{\l} Lis, Eshan Singhal, Nikola Mrk{\v{s}}i{\'c}, Tsung-Hsien Wen, and Ivan Vuli{\'c}.
\newblock Multilingual and cross-lingual intent detection from spoken data.
\newblock \emph{arXiv preprint arXiv:2104.08524}, 2021.

\bibitem[Goodfellow et~al.(2014)Goodfellow, Shlens, and Szegedy]{goodfellow2014explaining}
Ian~J Goodfellow, Jonathon Shlens, and Christian Szegedy.
\newblock Explaining and harnessing adversarial examples.
\newblock \emph{arXiv preprint arXiv:1412.6572}, 2014.

\bibitem[Huang et~al.(2024)Huang, Li, Yang, Shi, Chang, Ye, Wu, Hong, Huang, Liu, et~al.]{huang2024audiogpt}
Rongjie Huang, Mingze Li, Dongchao Yang, Jiatong Shi, Xuankai Chang, Zhenhui Ye, Yuning Wu, Zhiqing Hong, Jiawei Huang, Jinglin Liu, et~al.
\newblock Audiogpt: Understanding and generating speech, music, sound, and talking head.
\newblock In \emph{Proceedings of the AAAI Conference on Artificial Intelligence}, volume~38, pages 23802--23804, 2024.

\bibitem[Hughes et~al.(2025)Hughes, Price, Lynch, Schaeffer, Barez, Koyejo, Sleight, Perez, and Sharma]{hughesattacking}
John Hughes, Sara Price, Aengus Lynch, Rylan Schaeffer, Fazl Barez, Sanmi Koyejo, Henry Sleight, Ethan Perez, and Mrinank Sharma.
\newblock Attacking audio language models with best-of-n jailbreaking.
\newblock \emph{arXiv preprint}, 2025.

\bibitem[Jeong et~al.(2025)Jeong, Bae, Jung, Hwang, and Yang]{jeong2025playing}
Joonhyun Jeong, Seyun Bae, Yeonsung Jung, Jaeryong Hwang, and Eunho Yang.
\newblock Playing the fool: Jailbreaking llms and multimodal llms with out-of-distribution strategy.
\newblock In \emph{Proceedings of the Computer Vision and Pattern Recognition Conference}, pages 29937--29946, 2025.

\bibitem[Kang et~al.(2025)Kang, Xu, Yang, and Li]{kang2025advwave}
Mintong Kang, Chejian Xu, Shuang Yang, and Bo~Li.
\newblock Advwave: Stealthy adversarial jailbreak attack against large audio-language models.
\newblock In \emph{ICLR 2025 Workshop on Foundation Models in the Wild}, 2025.

\bibitem[Kingma(2014)]{kingma2014adam}
Diederik~P Kingma.
\newblock Adam: A method for stochastic optimization.
\newblock \emph{arXiv preprint arXiv:1412.6980}, 2014.

\bibitem[Lapid and Dubin(2025)]{lapid2025backdoors}
Raz Lapid and Almog Dubin.
\newblock Backdoors in conditional diffusion: Threats to responsible synthetic data pipelines.
\newblock \emph{arXiv preprint arXiv:2507.04726}, 2025.

\bibitem[Lapid and Sipper(2023)]{lapid2023see}
Raz Lapid and Moshe Sipper.
\newblock I see dead people: Gray-box adversarial attack on image-to-text models.
\newblock In \emph{Joint European Conference on Machine Learning and Knowledge Discovery in Databases}, pages 277--289. Springer, 2023.

\bibitem[Lapid et~al.(2022)Lapid, Haramaty, and Sipper]{lapid2022evolutionary}
Raz Lapid, Zvika Haramaty, and Moshe Sipper.
\newblock An evolutionary, gradient-free, query-efficient, black-box algorithm for generating adversarial instances in deep convolutional neural networks.
\newblock \emph{Algorithms}, 15\penalty0 (11):\penalty0 407, 2022.

\bibitem[Lapid et~al.(2024{\natexlab{a}})Lapid, Dubin, and Sipper]{lapid2024fortify}
Raz Lapid, Almog Dubin, and Moshe Sipper.
\newblock Fortify the guardian, not the treasure: Resilient adversarial detectors.
\newblock \emph{Mathematics}, 12\penalty0 (22):\penalty0 3451, 2024{\natexlab{a}}.

\bibitem[Lapid et~al.(2024{\natexlab{b}})Lapid, Langberg, and Sipper]{lapid2024open}
Raz Lapid, Ron Langberg, and Moshe Sipper.
\newblock Open sesame! universal black box jailbreaking of large language models.
\newblock In \emph{ICLR 2024 Workshop on Secure and Trustworthy Large Language Models}. ICLR, 2024{\natexlab{b}}.

\bibitem[Latif et~al.(2023)Latif, Shoukat, Shamshad, Usama, Ren, Cuay{\'a}huitl, Wang, Zhang, Togneri, Cambria, et~al.]{latif2023sparks}
Siddique Latif, Moazzam Shoukat, Fahad Shamshad, Muhammad Usama, Yi~Ren, Heriberto Cuay{\'a}huitl, Wenwu Wang, Xulong Zhang, Roberto Togneri, Erik Cambria, et~al.
\newblock Sparks of large audio models: A survey and outlook.
\newblock \emph{arXiv preprint arXiv:2308.12792}, 2023.

\bibitem[Li et~al.(2020)Li, Zhang, Jia, Xu, Zhang, Wang, Ma, and Gao]{li2020universal}
Jiguo Li, Xinfeng Zhang, Chuanmin Jia, Jizheng Xu, Li~Zhang, Yue Wang, Siwei Ma, and Wen Gao.
\newblock Universal adversarial perturbations generative network for speaker recognition.
\newblock \emph{arXiv preprint arXiv:2004.03428}, 2020.

\bibitem[Liu et~al.(2024)Liu, Xu, Chen, and Xiao]{liu2024autodan}
Xiaogeng Liu, Nan Xu, Muhao Chen, and Chaowei Xiao.
\newblock Auto{DAN}: Generating stealthy jailbreak prompts on aligned large language models.
\newblock In \emph{The Twelfth International Conference on Learning Representations}, 2024.
\newblock URL \url{https://openreview.net/forum?id=7Jwpw4qKkb}.

\bibitem[Madry et~al.(2018)Madry, Makelov, Schmidt, Tsipras, and Vladu]{madry2018towards}
Aleksander Madry, Aleksandar Makelov, Ludwig Schmidt, Dimitris Tsipras, and Adrian Vladu.
\newblock Towards deep learning models resistant to adversarial attacks.
\newblock In \emph{International Conference on Learning Representations}, 2018.

\bibitem[Mehrotra et~al.(2024)Mehrotra, Zampetakis, Kassianik, Nelson, Anderson, Singer, and Karbasi]{mehrotra2024tree}
Anay Mehrotra, Manolis Zampetakis, Paul Kassianik, Blaine Nelson, Hyrum Anderson, Yaron Singer, and Amin Karbasi.
\newblock Tree of attacks: Jailbreaking black-box llms automatically.
\newblock \emph{Advances in Neural Information Processing Systems}, 37:\penalty0 61065--61105, 2024.

\bibitem[Moosavi-Dezfooli et~al.(2017)Moosavi-Dezfooli, Fawzi, Fawzi, and Frossard]{moosavi2017universal}
Seyed-Mohsen Moosavi-Dezfooli, Alhussein Fawzi, Omar Fawzi, and Pascal Frossard.
\newblock Universal adversarial perturbations.
\newblock In \emph{Proceedings of the IEEE conference on computer vision and pattern recognition}, pages 1765--1773, 2017.

\bibitem[Neekhara et~al.(2019)Neekhara, Hussain, Pandey, Dubnov, McAuley, and Koushanfar]{neekhara2019universal}
Paarth Neekhara, Shehzeen Hussain, Prakhar Pandey, Shlomo Dubnov, Julian McAuley, and Farinaz Koushanfar.
\newblock Universal adversarial perturbations for speech recognition systems.
\newblock In \emph{Proc. Interspeech}, pages 481--485, 2019.

\bibitem[Nguyen et~al.(2025)Nguyen, Muller, Yu, Costa-Jussa, Elbayad, Popuri, Ropers, Duquenne, Algayres, Mavlyutov, et~al.]{nguyen2025spirit}
Tu~Anh Nguyen, Benjamin Muller, Bokai Yu, Marta~R Costa-Jussa, Maha Elbayad, Sravya Popuri, Christophe Ropers, Paul-Ambroise Duquenne, Robin Algayres, Ruslan Mavlyutov, et~al.
\newblock Spirit-lm: Interleaved spoken and written language model.
\newblock \emph{Transactions of the Association for Computational Linguistics}, 13:\penalty0 30--52, 2025.

\bibitem[Panayotov et~al.(2015)Panayotov, Chen, Povey, and Khudanpur]{panayotov2015librispeech}
Vassil Panayotov, Guoguo Chen, Daniel Povey, and Sanjeev Khudanpur.
\newblock Librispeech: an asr corpus based on public domain audio books.
\newblock In \emph{2015 IEEE international conference on acoustics, speech and signal processing (ICASSP)}, pages 5206--5210. IEEE, 2015.

\bibitem[Qin et~al.(2019)Qin, Carlini, Cottrell, Goodfellow, and Raffel]{qin2019imperceptible}
Yao Qin, Nicholas Carlini, Garrison Cottrell, Ian Goodfellow, and Colin Raffel.
\newblock Imperceptible, robust, and targeted adversarial examples for automatic speech recognition.
\newblock In \emph{International conference on machine learning}, pages 5231--5240. PMLR, 2019.

\bibitem[Radford et~al.(2023)Radford, Kim, Xu, Brockman, McLeavey, and Sutskever]{radford2023robust}
Alec Radford, Jong~Wook Kim, Tao Xu, Greg Brockman, Christine McLeavey, and Ilya Sutskever.
\newblock Robust speech recognition via large-scale weak supervision.
\newblock In \emph{International conference on machine learning}, pages 28492--28518. PMLR, 2023.

\bibitem[Rubenstein et~al.(2023)Rubenstein, Asawaroengchai, Nguyen, Bapna, Borsos, Quitry, Chen, Badawy, Han, Kharitonov, et~al.]{rubenstein2023audiopalm}
Paul~K Rubenstein, Chulayuth Asawaroengchai, Duc~Dung Nguyen, Ankur Bapna, Zal{\'a}n Borsos, F{\'e}lix de~Chaumont Quitry, Peter Chen, Dalia~El Badawy, Wei Han, Eugene Kharitonov, et~al.
\newblock Audiopalm: A large language model that can speak and listen.
\newblock \emph{arXiv preprint arXiv:2306.12925}, 2023.

\bibitem[Sadasivan et~al.(2025)Sadasivan, Feizi, Mathews, and Wang]{sadasivan2025attackersnoise}
Vinu~Sankar Sadasivan, Soheil Feizi, Rajiv Mathews, and Lun Wang.
\newblock Attacker's noise can manipulate your audio-based llm in the real world.
\newblock \emph{arXiv preprint arXiv:2507.06256}, 2025.

\bibitem[Saon et~al.(2025)Saon, Dekel, Brooks, Nagano, Daniels, Satt, Mittal, Kingsbury, Haws, Morais, et~al.]{saon2025granite}
George Saon, Avihu Dekel, Alexander Brooks, Tohru Nagano, Abraham Daniels, Aharon Satt, Ashish Mittal, Brian Kingsbury, David Haws, Edmilson Morais, et~al.
\newblock Granite-speech: open-source speech-aware llms with strong english asr capabilities.
\newblock \emph{arXiv preprint arXiv:2505.08699}, 2025.

\bibitem[Shen et~al.(2024)Shen, Wu, Backes, and Zhang]{shen2024voicejailbreak}
Xinyue Shen, Yixin Wu, Michael Backes, and Yang Zhang.
\newblock Voice jailbreak attacks against gpt-4o.
\newblock \emph{arXiv preprint arXiv:2405.19103}, 2024.

\bibitem[Tamam et~al.(2023)Tamam, Lapid, and Sipper]{vitrack-tamam2023foiling}
Snir~Vitrack Tamam, Raz Lapid, and Moshe Sipper.
\newblock Foiling explanations in deep neural networks.
\newblock \emph{Transactions on Machine Learning Research}, 2023.
\newblock ISSN 2835-8856.
\newblock URL \url{https://openreview.net/forum?id=wvLQMHtyLk}.

\bibitem[Tang et~al.(2024)Tang, Yu, Sun, Chen, Tan, Li, Lu, MA, and Zhang]{tang2024salmonn}
Changli Tang, Wenyi Yu, Guangzhi Sun, Xianzhao Chen, Tian Tan, Wei Li, Lu~Lu, Zejun MA, and Chao Zhang.
\newblock {SALMONN}: Towards generic hearing abilities for large language models.
\newblock In \emph{The Twelfth International Conference on Learning Representations}, 2024.
\newblock URL \url{https://openreview.net/forum?id=14rn7HpKVk}.

\bibitem[Vadillo and Santana(2022)]{vadillo2022human}
Jon Vadillo and Roberto Santana.
\newblock On the human evaluation of universal audio adversarial perturbations.
\newblock \emph{Computers \& Security}, 112:\penalty0 102495, 2022.

\bibitem[Warden(2018)]{warden2018speech}
Pete Warden.
\newblock Speech commands: A dataset for limited-vocabulary speech recognition.
\newblock \emph{arXiv preprint arXiv:1804.03209}, 2018.

\bibitem[Xie and Wu(2024)]{xie2024mini}
Zhifei Xie and Changqiao Wu.
\newblock Mini-omni2: Towards open-source gpt-4o with vision, speech and duplex capabilities.
\newblock \emph{arXiv preprint arXiv:2410.11190}, 2024.

\bibitem[Zhang et~al.(2023)Zhang, Li, Zhang, Zhan, Wang, Zhou, and Qiu]{zhang2023speechgpt}
Dong Zhang, Shimin Li, Xin Zhang, Jun Zhan, Pengyu Wang, Yaqian Zhou, and Xipeng Qiu.
\newblock Speechgpt: Empowering large language models with intrinsic cross-modal conversational abilities.
\newblock In \emph{Findings of the Association for Computational Linguistics: EMNLP 2023}, pages 15757--15773, 2023.

\bibitem[Zhang et~al.(2024)Zhang, Ye, Ma, Li, Yang, Sang, and Yeung]{zhang2024anyattack}
Jiaming Zhang, Junhong Ye, Xingjun Ma, Yige Li, Yunfan Yang, Jitao Sang, and Dit-Yan Yeung.
\newblock Anyattack: Towards large-scale self-supervised generation of targeted adversarial examples for vision-language models.
\newblock \emph{arXiv e-prints}, pages arXiv--2410, 2024.

\bibitem[Zhou et~al.(2024)Zhou, Wang, Xiong, Xia, Gu, Chai, Zhu, Huang, Dou, Xi, et~al.]{zhou2024easyjailbreak}
Weikang Zhou, Xiao Wang, Limao Xiong, Han Xia, Yingshuang Gu, Mingxu Chai, Fukang Zhu, Caishuang Huang, Shihan Dou, Zhiheng Xi, et~al.
\newblock Easyjailbreak: A unified framework for jailbreaking large language models.
\newblock \emph{arXiv preprint arXiv:2403.12171}, 2024.

\bibitem[Zhou et~al.(2025)Zhou, Lou, Huang, Qin, Yang, and Wang]{zhou2025don}
Yukai Zhou, Jian Lou, Zhijie Huang, Zhan Qin, Sibei Yang, and Wenjie Wang.
\newblock Don’t say no: Jailbreaking llm by suppressing refusal.
\newblock In \emph{Findings of the Association for Computational Linguistics: ACL 2025}, pages 25224--25249, 2025.

\bibitem[Zou et~al.(2023)Zou, Wang, Carlini, Nasr, Kolter, and Fredrikson]{zou2023universal}
Andy Zou, Zifan Wang, Nicholas Carlini, Milad Nasr, J~Zico Kolter, and Matt Fredrikson.
\newblock Universal and transferable adversarial attacks on aligned language models.
\newblock \emph{arXiv preprint arXiv:2307.15043}, 2023.

\end{thebibliography}

\end{document}